\RequirePackage{amsmath}
\documentclass[runningheads]{llncs}

\usepackage{cite}
\usepackage{amssymb,amsfonts} 
\usepackage{graphicx} % Additional graphics capabilities
\usepackage{wasysym} % Symbols for attacker profile partially ordered sets
\usepackage{csvsimple} % Creates tables from .csv files
\usepackage{paralist} % In-paragraph lists

\graphicspath{ {./images/} }
\begin{document}
\title{An Attacker Modeling Framework for the Assessment of Cyber-Physical Systems Security}
\titlerunning{CPS Attacker Modeling Framework}

\author{Christopher Deloglos \and Carl Elks \and Ashraf Tantawy}
\authorrunning{C. Deloglos et al.}
\institute{Virginia Commonwealth University, Richmond VA 23220, USA
\email{delogloscj,crelks,amatantawy@vcu.edu}}
\maketitle

\begin{abstract}
\begin{sloppypar}
Characterizing attacker behavior with respect to Cyber-Physical Systems is important to assuring the security posture and resilience of these systems. Classical cyber vulnerability assessment approaches rely on the knowledge and experience of cyber-security experts to conduct security analyses and can be inconsistent where the experts' knowledge and experience are lacking. This paper proposes a flexible attacker modeling framework that aids in the security analysis process by simulating a diverse set of attacker behaviors to predict attack progression and provide consistent system vulnerability analysis. The model proposes an expanded architecture of vulnerability databases to maximize its effectiveness and consistency in detecting CPS vulnerabilities while being compatible with existing vulnerability databases. The model has the power to be implemented and simulated against an actual or virtual CPS. Execution of the attacker model is demonstrated against a simulated industrial control system architecture, resulting in a probabilistic prediction of attacker behavior.
\end{sloppypar}
\end{abstract}


\section{Introduction}
% Problem Description
To secure systems from known and emerging threats, systems engineers and security analysts alike need to integrate an attacker's view of potential vulnerabilities into their design, development, and analysis process as early as possible.
To date, this attacker perspective activity has been largely a manual process conducted by subject matter experts who examine a system and identify possible vulnerabilities and weaknesses. Understanding the potential threat behaviors and capabilities of a cyber-actor or threat agent with respect to a cyber or cyber-physical system (CPS) is critical for risk assessment and the development of effective security countermeasure solutions.~\cite{Knowles2015b}~\cite{Mo2012Cyber-physicalInfrastructure}.

% Importance of the Problem
A promising way forward is the use of formal attacker models that attempt to characterize and capture the expected behavior of attackers against a CPS. Developing accurate attacker-models, however, is difficult due to the number of factors that influence and encompass attacker behavior.
To understand the behavior of an attacker many questions may be asked such as ``Who is the attacker?", ``What resources are at the attacker's disposal?", and ``What is the motivation of the attacker?". Quantifying the answers to these questions, however, is challenging in the absence of absolute metrics and incomplete situation awareness. To partially address these challenges, in this paper we present a novel modular attacker modelling framework (AMF) that is: (1) capable of capturing and quantifying complex attacker behavior and, (2) is readily extensible to incorporate new or emerging aspects of attacker behavior. 

\section{Related work}
% Attacker Profiles
In attacker modeling, a common approach is to create a correlation model where the designer selects a series of attacker properties such as skill level, resources, intent, and motivation, and attempts to develop cumulative correlation functions that effectively predict attacker behavior when applied to real-world attackers~\cite{Heckman2005}. Rocchetto et al. performed a literature search and created a six-profile model able to effectively describe attacker profiles from the majority of cited literature~\cite{Rocchetto2016}. The use of this method, however, requires the user to be aware of the skill level of the attacker. In order to apply Rocchetto's method to a realistic attacker model, a probability mass function is used to simulate the non-determinism in the skillset of an unknown attacker.

% Modeling attacker behavior
Orojloo et al.~\cite{Orojloo2016PredictingSystems} proposed an attack modeling approach that applied attack trees to model how characteristics of a particular attack (access, knowledge required, skill required, and level of user interaction) influence the behavior of an attacker. While serving as an effective foundation, the attack tree method as proposed by Orojloo requires a comprehensive perspective of the CPS which must be created through a manual process for each CPS under consideration. This design process becomes tedious when considering a large CPS or one with multiple attack vectors to a single target.

% Markov Modeling
A common approach to representing attacker behavior against a CPS is application of the Markov decision process (MDP)~\cite{Kriaa2012, Chen2018}. The size of the Markov model explodes, however, when considering a large CPS with multiple nodes, each having multiple potential actions. Markov representation also lacks the ability to clearly depict the nature of various paths an attacker may take through a system. An alternate to the full MDP is the Partially Observable Markov Decision Process (POMDP) as in~\cite{Carin2008QuantitativeMethodology}. This model allows the application of the Markov methodology while limiting the state-space of the model to a single attack-path. We propose an alternate scheme using a one-step look-ahead formalism as a solution.

% Application of online databases to attack-process
Automation of the attack-analysis process for a complex CPS was explored in~\cite{LeMay2011}, where the ADVISE method is proposed. The ADVISE method requires as input a description of the system, a description of the adversary, a list of the desired security metrics, and a description of all vulnerability information pertaining to the system. In application to a real system, researching, compiling, and organizing all vulnerability information related to a CPS is a monumental task. Databases such as the Common Attack Pattern Enumeration and Classification (CAPEC), the Common Weakness Enumeration (CWE), the Common Vulnerabilities and Exposures (CVE), and the Common Platform Enumeration (CPE) have been applied to attack modeling to aid in vulnerability research~\cite{Mili2019, Ekelhart2015}. We propose the application of CAPEC, CWE, CVE, and CPE search engines as an aid for the generation of a hybrid action database.

% Universal modeling attempts
Due to the diversity of CPSs, attacker models are often defined within the context of a specific system~\cite{Mili2019, Ekelhart2015, Adepu2016}. In this paper, we propose a modular AMF that is capable of modeling complex attacker behavior against a full CPS and is readily expandable to include additional aspects of attacker behavior. The authors propose two components of the attacker model as novel. First, the authors propose a formalization between an attacker profile and an attacker's behavior that allows the prediction of the behavior of profile-specific attackers against a CPS. This approach allows an attacker-specific security review and aids in prioritizing the relevance of attacks and vulnerabilities in a CPS. Second, the authors propose a modular scheme for describing cyber-physical system architectures as well as the attack progression of an attacker through cyber-physical systems.
\par
The remainder of the paper is organized as follows. Section \ref{sect:overview} describes the attacker-modeling framework. A case study is explored in section \ref{sect:case study}. Section \ref{sect:conclusion} summarizes the work and proposes future research directions.

\section{Overview of the Attacker Model} \label{sect:overview}
\begin{figure}
    \centering
    \includegraphics[width=\textwidth]{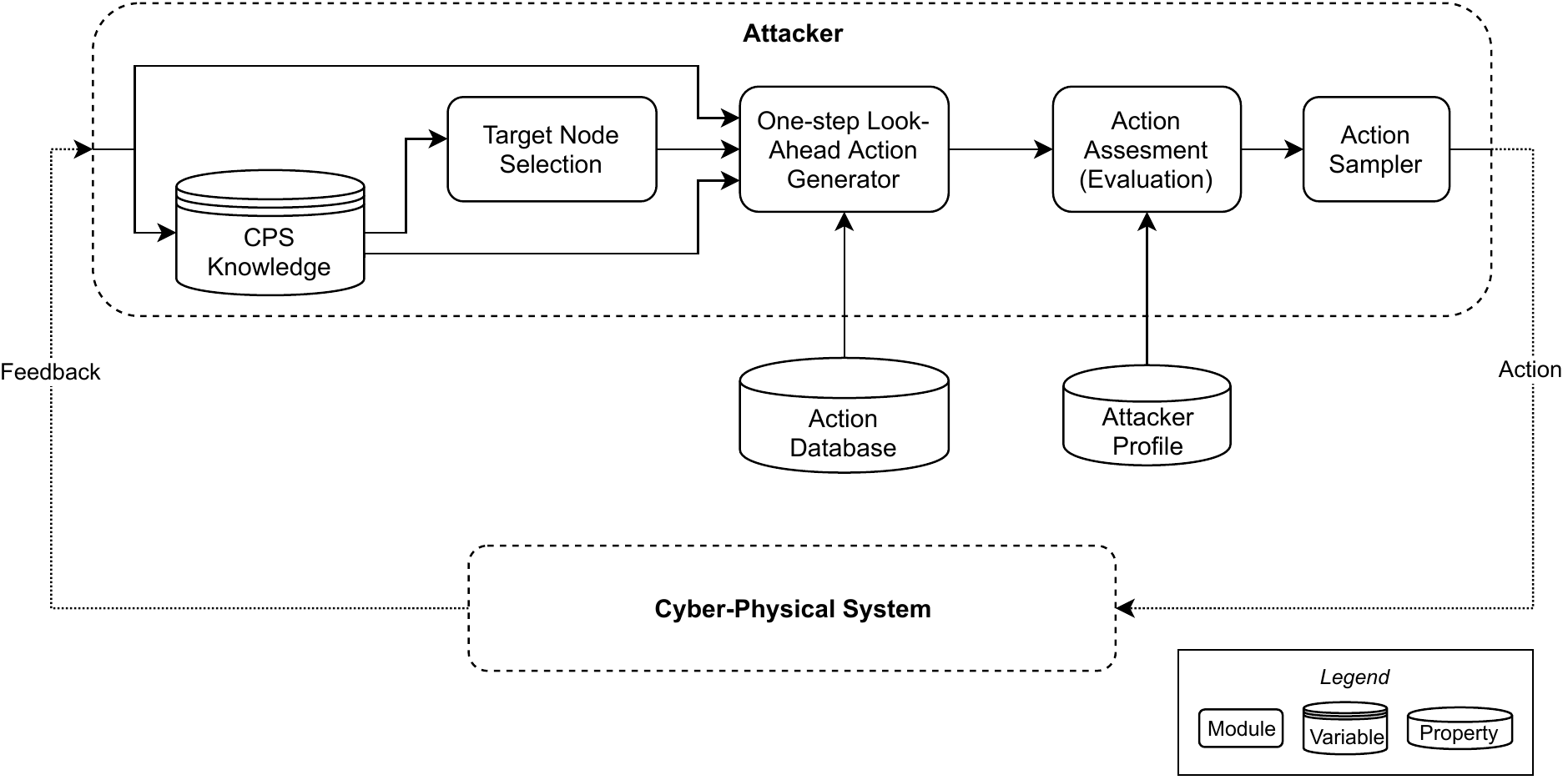}
    \caption{Attacker Model Overview}
    \label{fig:Attacker Model Overview}
\end{figure}

Traditional attacker behavior is captured as a series of observations regarding the attacker motivation and decision process. These observations are captured in the proposed AMF as \textit{rules}. A rule may be formally defined as a facet of an attacker's behavior which observes a cause/effect relationship between an influencing parameter and the attacker's actions. The proposed AMF accepts a series of rules which together define the behavior of the attacker.
The attacker model overview in Figure~\ref{fig:Attacker Model Overview} depicts the relationship between the attacker and the CPS. Sample rules defining attacker behavior are implemented in modules such as the CPS Knowledge and the Target Node Selection modules. This provides a test-bed to explore the role and influence of individual rules on the composite attacker decision process. This also provides a flexible framework that allows validation of the AMF against a particular dataset where the rules the AMF implements may be refined and calibrated to achieve a model that accurately reflects the behavior of a known attacker or set of attackers.
\par
Relationships between modules are characterized by connecting arrows, where upstream modules occur earlier in the attacker decision process.
The proposed attacker model operates on a cyclical action/feedback scheme where an \textit{action} is some step the attacker performs in the attack process and \textit{feedback} is any information the attacker receives as a result of the action.
The progression of the attack through the CPS is described as the \textit{attack state}.
The attack state represents a single action and is a tuple of static \textit{properties} and dynamic \textit{variables}.
Properties are information that is static throughout the attack process and in Fig.~\ref{fig:Attacker Model Overview} includes the Action Database and the Attacker Profile.
Variables include information that changes as the attack progresses such as the attacker CPS Knowledge.
The action performed for each state of the attack is a product of all static properties and dynamic variables and is discussed further in section \ref{sect:Action Assessment}.

\subsection{Cyber-Physical System} \label{sect:CPS}
In the proposed attacker-modeling framework, the CPS is modeled and described as a composition of \textit{nodes}, \textit{edges}, \textit{attack vectors}, and \textit{entry points}. A node represents a machine or other potentially vulnerable device that has functional purpose within the CPS. An edge represents a communication link between two nodes that may be used to transmit information, while an attack vector is any edge that may be used for an attack. An entry point is an edge directed into the CPS from outside the CPS that may be used by an attacker to gain access to the system. Establishing well-formed boundaries between nodes of the CPS and relationships between them allows a formal description of a Cyber-Physical System. Programming tools such as SysML~\cite{SysMLProject} utilize the node/edge system description scheme and may be readily integrated with the CPS design process for attacker model automation.\par

\subsection{CPS Knowledge} \label{sect:CPSK}
We begin with making the reasonable assumption that as the attack or probing progresses, the attacker will begin to learn information about the target system. This behavior is captured in the CPS knowledge module. The rules that the CPS knowledge module implements are:
\begin{itemize}
    \item When starting an attack, the attacker only has knowledge of the system entry points.
    \item As the attack progresses the attacker will discover new information about the CPS.
    \item If a node is compromised, all nodes it is connected to are discovered and added to the CPS knowledge.
\end{itemize}
Information about the CPS is fed into the attacker's CPS knowledge module as feedback. If a node is compromised it is considered owned by the attacker and capable of performing pivoting attacks. The initial attacker CPS knowledge is simplified for demonstration purposes, but in a more complex application could include behaviors involving the attacker's discovery of the target system. More complex rules may be applied to node ownership such as defining levels of ownership (based on privilege escalation).

\subsection{Target Node Selection}
When the attacker goes to perform an action against a system, the attacker must first select a target node. The rules that the target node selection module implements are:
\begin{itemize}
    \item The attacker will only target nodes that exist in the attacker's CPS Knowledge.
    \item The attacker will not target a node if it is already compromised.
    \item The attacker will not target a node if the attacker has exhausted all qualified actions against it.
    \item If an attacker targets a node, the attacker will not change targets until exhausting all actions against it.
    \item If more than one target is valid, the attacker will select a target node at random from amongst the valid nodes.
\end{itemize}
Several additional rules may be added to capture the tendency of an attacker to target nodes associated with the end-goal (often referred to as the honey-pot).

\subsection{Action Database}
The purpose of the Action Database module is to capture behaviors influenced by the actions available to the attacker. The action database contains descriptions of all actions known about the CPS. Each action within the database contains several fields of information including the action profile, the action description, the target criteria, and a list of prerequisite actions. The \textit{action profile} contains a quantitative description of the user and use-case of the action, which is used for quantifying a relationship between each action and the attacker profile defined for the attacker model as discussed further in section \ref{sect:attack profiles}. The \textit{action description} is a plain-text description of how the action works in as much detail as is possible. The \textit{target criteria} defines what system(s) the action is valid against. The \textit{prerequisite attacks} describe any actions that must be completed before this action may be attempted.
\par
A critical component to the viability of the attacker model is the database population scheme. CAPEC~\cite{MITRECorporation2011}, CWE\cite{2018CWEEnumeration}, CVE\cite{MITRE2016}, and CPE~\cite{2019CPEEnumeration} are amongst the most popular and provide different approaches to cataloguing attacks, attack descriptions, and attack relationships. Search engines that make use of online attack and vulnerability databases aid in effectively generating an action database for the attacker model. One tool that was applied to populate the action database for the case study in section \ref{sect:case study} was the CYBOK tool~\cite{8850328}, which is a literal search engine for CAPEC, CWE, and CVE capable of generating vulnerability data for individual queries or entire systems.

\subsection{One-Step Look-Ahead Generator} \label{sect:OSLAG}
\begin{figure}
	\centering
	\includegraphics[width=6cm]{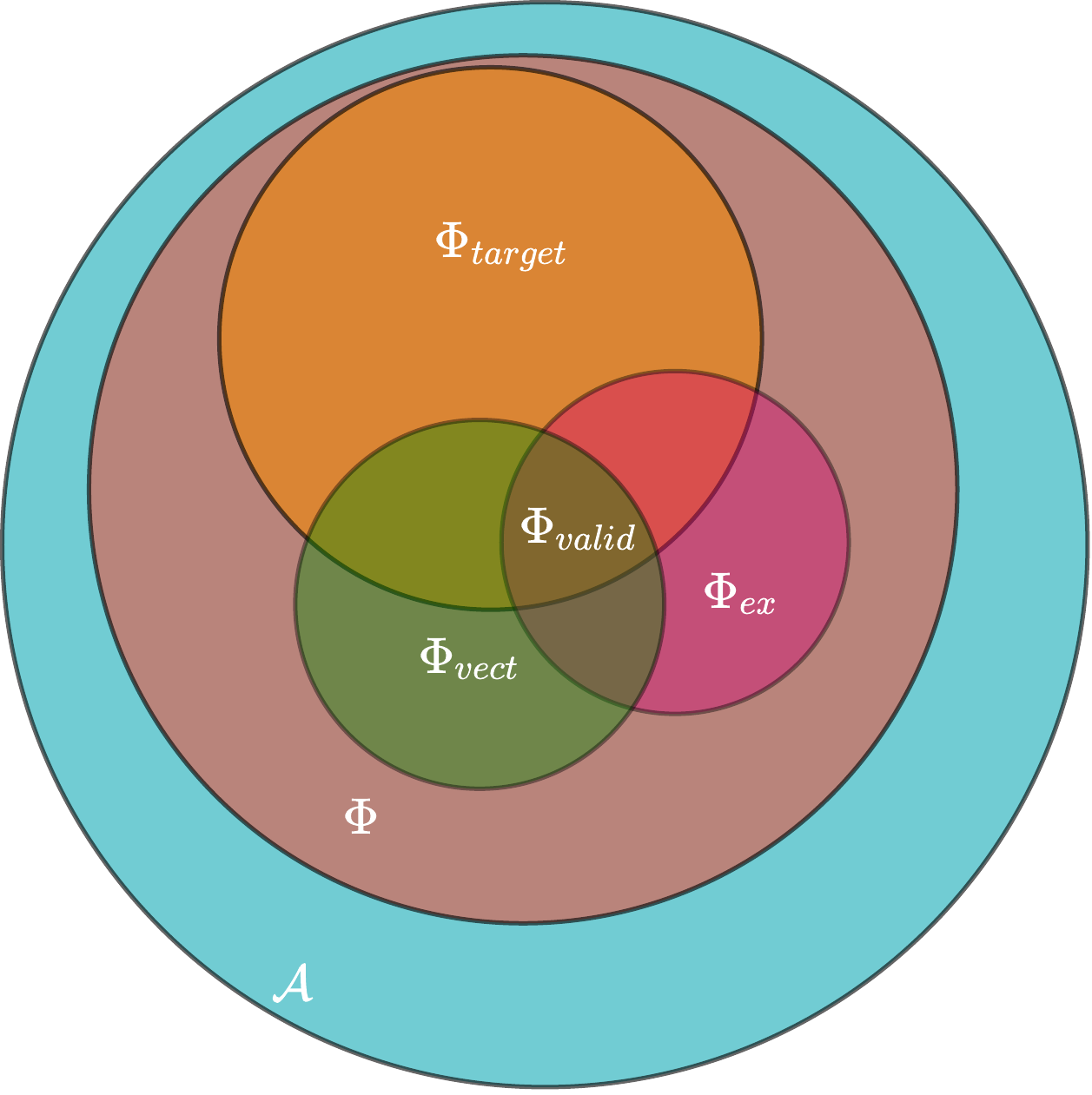}
	\caption{The intersection of action selection filters applied to the action database} 
	\label{fig:targetnodeselection}
\end{figure}

The one-step look-ahead generator applies the attacker's knowledge of the CPS to filter out all attacks that are invalid for the current attack state. Filters are non-probabilistic in nature and may depend on any information regarding the current state of the attack or the description of the node. This attacker model applies three filters.
\begin{enumerate}
    \item The attacker will only consider actions that meet the target criteria
    \item The attacker will not consider an action that has already been performed on the target
    \item The attacker will not consider an action if the edge relating the current node to the target node is not a viable propagation path for that action
\end{enumerate}
We define $\mathcal{A}$ as the set of all $m$ known actions in the action database and $\Phi \subseteq \mathcal{A}$ as the set of actions known by the attacker. The three filters are defined as $\Phi_{target} \subseteq \Phi$, $\Phi_{ex} \subseteq \Phi$, and $\Phi_{vect} \subseteq \Phi$ for filters 1, 2, and 3 respectively. The set of actions that are valid for the attacker to perform in the given state of the attack ($\Phi_{valid}$) are then defined by Equation \ref{eq:Valid actions}, where the attack space can be visualized in Fig.~\ref{fig:targetnodeselection}.
\begin{equation} \label{eq:Valid actions}
    \Phi_{valid} = \Phi_{ex} \cap \Phi_{vect} \cap \Phi_{target}
\end{equation}

\subsection{Action Assessment} \label{sect:Action Assessment}
When an attacker goes to select an action, characteristics about the attacker will influence the selection of the attack. This behavior is captured in the action assessment module. The rules applied here are:
\begin{itemize}
    \item An attacker's behavior is dependent on one or more primary influencing factors.
    \item Actions may have properties that allow them to be correlated to attackers.
    \item An attacker's attack selection decision can be predicted by evaluating the sum of influencing factors between an attacker and an action.
\end{itemize}
The action assessment module calculates the probability of the attacker performing each of the actions based on probability functions that take as operands the attacker profile, the attack profile, and the current state of the attack.

\subsubsection{Attacker Profiles and Attack Profiles} \label{sect:attack profiles}
\textit{Attacker profiles} are a topic well covered in literature with no recognized standards for what characteristics best model an attacker. The purpose of an attacker profile is to capture characteristics about an attacker that influence the attacker's behavior, thereby describing the expected behavior of the attacker. The characteristics that define the attacker profile are termed attacker \textit{properties}. Rocchetto et al.~\cite{Rocchetto2016} performed a literature review on attacker profiles for CPSs in an attempt to find a unifying attacker profiling model to describe various attackers from multiple different research studies. In conclusion, Rocchetto proposed a set of six attacker profiles composed of twenty-nine attacker properties that effectively described the majority of attacker profiles in the referenced literature. 

An \textit{action profile} is often represented as a set of properties describing the characteristics of the action~\cite{Rocchetto2016}. This, however, implies linearly proportionality to an attacker profile, which is not universally true. For example, an attacker with a high skill set is not necessarily more likely to perform an attack that requires a high skill set when an easier attack may succeed as well. We capture this behavior by defining an attack profile as the profile of the attacker expected to use that attack. Because attacker behavior is constantly changing as technology evolves, this profiling technique may be reinforced by empirical data from records of attack history. Collaborations such as MITRE's ATT\&CK framework \cite{2020ATTCKSystems} may aid in assessment of current threat actors. This facilitates an attacker model that can better emulate realistic and relevant threats by allowing the user to base the relationship between attackers and their actions off of current attacker data.
\par
\begin {figure}
\centering
\begin{minipage}[b]{.48\textwidth}
    \centering
    \includegraphics[width=\textwidth]{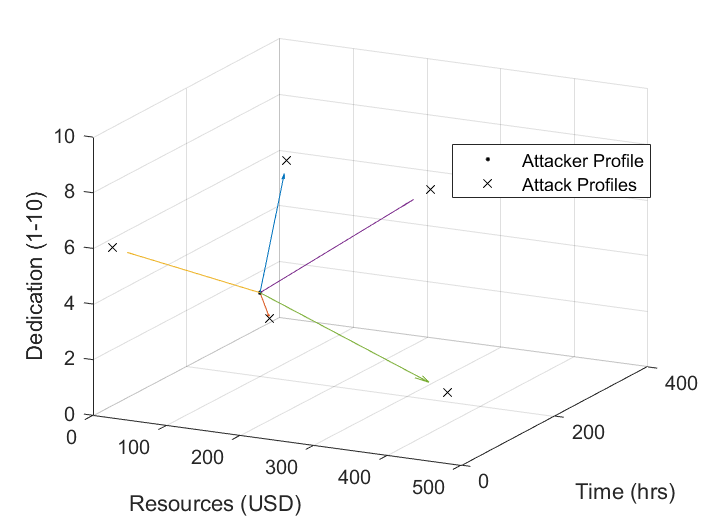}
    \caption{An example 3-dimensional attack space showing the attacker profile and several action profiles}
    \label{fig:3dattackproperties}
\end{minipage}
\hfill
\begin{minipage}[b]{.48\textwidth}
    \centering
    \includegraphics[width=\textwidth]{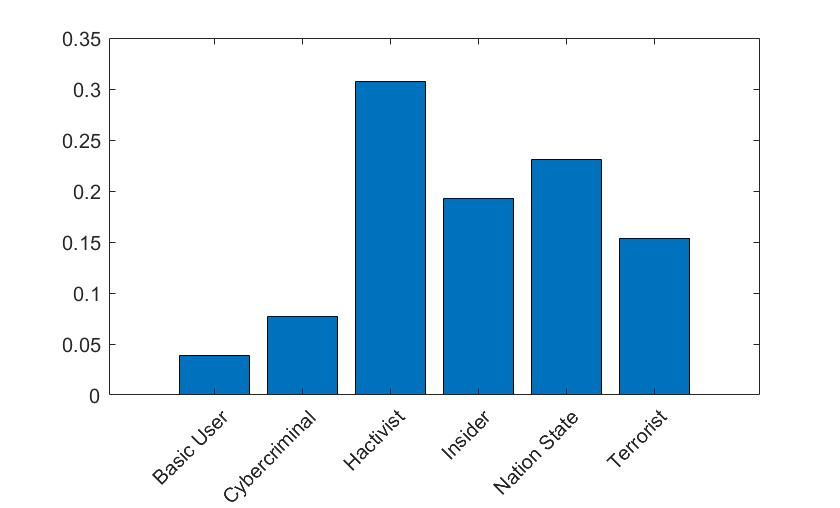}
	\caption{Example probability mass function for a probabilistic attacker profile against a nuclear power plant}
	\label{fig:examplePDF}
\end{minipage}
\end{figure}

As such, we define an attacker profile ($\Delta$) as an $n$-dimensional space of attacker properties ($\delta_i$) such that $\Delta=\{\delta_1, \delta_2,\ldots, \delta_n\}$ for an attacker profile having $n$ properties. An example attack space can be seen in Fig.~\ref{fig:3dattackproperties} where the attacker profile and several action profiles are plotted in the 3-dimensional space. The probability of an attacker performing an action is a function of the distance between the attacker profile and the action profile in n-dimensional space.

\subsubsection{Probabilistic Attacker Profiles} \label{sect:probabilisticattackerprofiles}
In applying Rocchetto's attacker profiles to an attacker model it is important to note that in a real-world application one cannot assume which attacker will be attacking a system. In order to simulate this non-deterministic behavior, two types of attacker profiles are adopted which are the \textit{static attacker profile} and the \textit{probabilistic attacker profile}. A static attacker profile represents one of the six attacker profiles defined by Rocchetto et. al. \cite{Rocchetto2016}. A probabilistic attacker profile may be represented as a probability mass function (PMF) of the six profiles. The PMF is generated by assigning each of the six attacker profiles ($\Delta_1,\ldots,\Delta_6$) a likelihood of attacking ($l_i$) such that $0\leq l_i\leq 1$.The Probability of attack of a specific attacker profile is calculated using:
\begin{equation}   \label{eq:PDF for NPP}
    P(\Delta_i) = \frac{l_i}{\sum_{j=1}^n l_j}
\end{equation}
where $\sum_{j=1}^nP(\Delta_j)=1$. The PMF in Fig.~\ref{fig:examplePDF} is an example probabilistic attacker profile designed to mimic the probability of attackers against a nuclear power plant. At the beginning of the attack analysis process, the probabilistic attacker profile is sampled to obtain a discrete attacker profile which is recognized as the attacker for the remainder of the attack process.

\subsubsection{Attack Probability Functions} \label{sect:Action Probability Function}
The probability that the attacker will perform an attack at any given time is calculated using the attack probability function. Attacker properties may be one of three types which are sets, bounded ranges, and unbounded ranges. Non-ordered sets are considered to have a scaled property value $\gamma=1$ if the attacker profile property and the attack profile property match and $\gamma=0$ otherwise. Ordered set values may be mapped to the scaled property range ($0\leq \gamma\leq 1$) using Fuzzy set theory as demonstrated by Patil et al. in~\cite{Patil2014}.
\par
Bounded ranges are numerical ranges where the value of a property ($\varepsilon$) may only fall between a lower bound ($\varepsilon_L$) and an upper bound ($\varepsilon_H$). Bounded ranges are linearly mapped to the scaled property value ($\gamma: 0\leq\gamma\leq1$) using:
\begin{equation}
    \gamma = \frac{\varepsilon-\varepsilon_L}{\varepsilon_H-\varepsilon_L}
    \label{eq:1}
\end{equation}
\par
Several scaling functions exist for unbounded ranges such as the percent-difference function, the logistic function, and the hyperbolic tangent. The value weighting in these functions, however, is non-linear, which does not properly scale different property values where a score considered median is represented by a numerically large or numerically small value ($>$100 or $<$1 respectively). Therefore, we propose converting the unbounded property values to a bounded range by first evaluating the maximum ($\gamma_{max}$) and minimum ($\gamma_{min}$) values for all actions within the database, then using the local maximum and minimum to scale the unbounded range.

We designate the set of $m$ available actions in the action database as $\mathcal{A}=\{A_1,A_2,\ldots,A_m\}$. Each action $A_i$ ($i=1,2,\ldots, m$) has an associated set of scaled property values $\Gamma_i=\{\gamma^1_i, \gamma^2_i,\ldots, \gamma^n_i\}$. For a given attacker profile ($\Delta$) with $n$ scaled property values $\Theta=\{\theta_1, \theta_2, \ldots, \theta_n\}$, the distance ($d_i$) between the attacker and each action is calculated by the distance between the two profiles in $n$-dimensional space using:
% Using Modified Mahalanobis distance equation
\begin{equation}
    d_i = f(\Theta;\Gamma_i) =\sqrt{\sum_{j=1}^n \frac{1}{\beta_j^2}\left(\theta_j - \gamma_i^j\right)^2}
    \label{eq:2}
\end{equation}

where $\beta_j$ is a criticality factor such that $\{\beta\in\mathbb{R}|0\le\beta\le1\}$ which increases the distance for properties with a $\beta<1$ criticality. The score of each action ($s_i$) is inversely proportional to $d_i$ and calculated using the function:
\begin{equation} \label{eq:scoring}
    s_i = 1-\frac{d_i}{\sum_{j=1}^md_j} \qquad i=1,\ldots,m
\end{equation}
This equation is unique in that it calculates the inverse of the distance without applying a nonlinear value-weighting as is observed in the inverse function or exponential functions such as the Softmax function. According to the score for each action, the probability that the attacker will take action $A_i$ is calculated using the function:
\begin{equation} \label{eq:Prob_Assignment}
    P[A_i] = \frac{s_i}{\sum_{j=1}^ms_j}
\end{equation}
Equation (\ref{eq:Prob_Assignment}) has the intuitive interpretation that the higher the score the attacker gets for an action, the higher the probability that this action will be chosen by the attacker.

\subsection{Action Sampler} \label{sect:sampler}
The last module in the attacker model is the action sampler module. The action sampler module implements the following rule: After evaluating the attacks, the attacker is more likely to choose an attack with a high probability than an attack with a low probability.

The action sampler receives as inputs all attacks for the target system with their attack probability values and selects one of the actions by sampling a weighted randomizing function ($rand_w()$) mapped to the probabilities of the set of probabilistic actions $\Delta=\{A_i,P[A_i]\}$. This action is then performed by the attacker against the CPS.

\section{Case Study} \label{sect:case study}
\begin{figure}
	\centering
	\includegraphics[width=\textwidth]{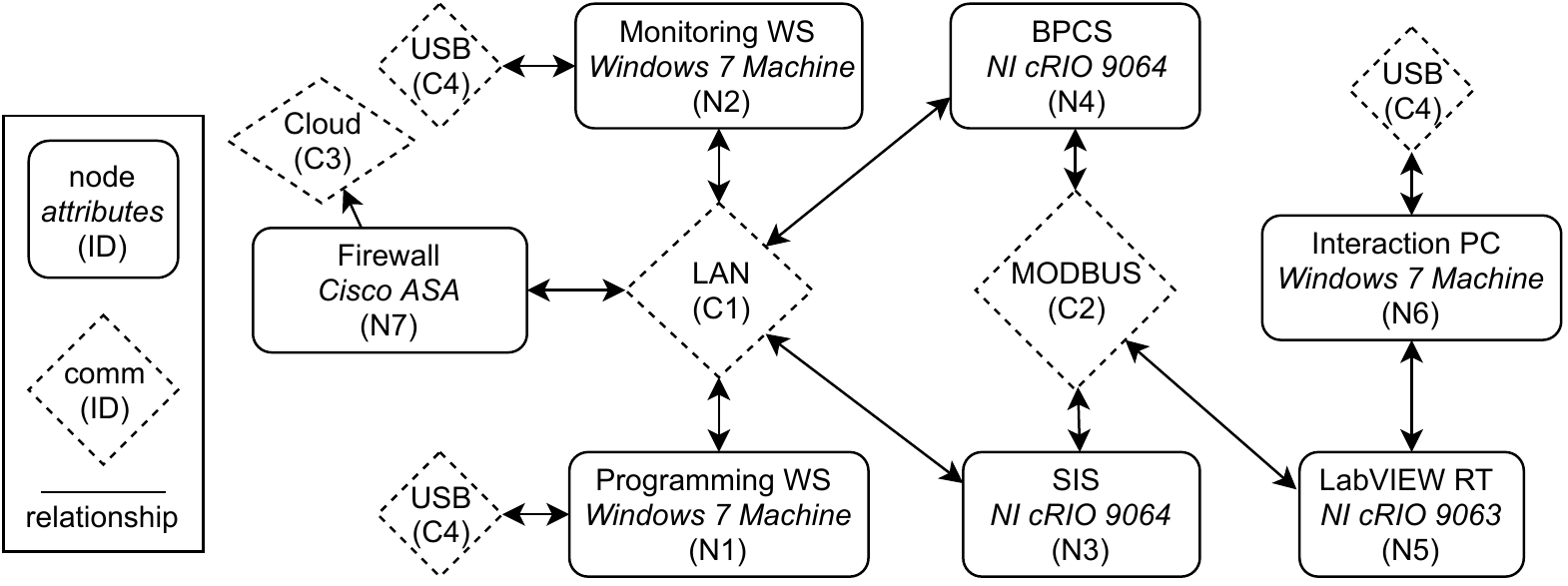}
	\caption{Case Study ICS relational diagram}
	\label{fig:casestudy1}
\end{figure}

For application of the proposed AMF, we observe the Industrial Control System (ICS) in Fig.~\ref{fig:casestudy1} composed of nodes and communication channels. This example ICS is used to control a simulated exothermic continuous stirred tank reactor (CSTR) using an NI cRIO controller. The target for the attack is the Basic Process Control System (BPCS, N4). Control or disruption of the BPCS by the attacker indicates a successful attack.

\subsection{ICS Formal Description}
The ICS consists of 7 nodes, each composed of key attributes included in Fig.~\ref{fig:casestudy1}. The system is described as having 4 entry points which include N1, N2, and N6 via infected USB and N7 via remote access. Six properties are selected as a subset of those described by Rocchetto et al.~\cite{Rocchetto2016} to describe the attacker and action profiles which include Access, Finances, Knowledge, Manpower, Motivation, and Tools. Access, Motivation, and Tools are defined as set properties with values of \{Direct, Wireless, Offsite\} for Access and \{Low, Medium, High\} for Motivation and for Tools. Knowledge is defined as a bounded property with a {0 $\leq$ Knowledge $\leq$ 10} range. Finances and Manpower are defined as unbounded properties. These properties are not intended to be a holistic description of the attacker behavior, but rather to demonstrate the principles and dynamics of the different types of profile properties. The criticality factor is kept at unity (1) for all profile properties. The attacker profile PMF in Fig.~\ref{fig:examplePDF} was defined as a set of 6 attacker profiles with property values in Table~\ref{table:probabilistic attacker profile}. CAPEC, CWE, CVE, and CPE databases were used to search for vulnerability information. The CAPEC and CWE databases were used to identify potential attack patterns and weaknesses respectively, aiding in the discovery of associated CVEs. Table~\ref{table:cs vulnerabilities values} contains a sample profile set for the vulnerabilities found for the ICS nodes.

\begin{table}
    \caption{Attacker profiles and property values}
	\centering
	\begin{filecontents*}{data/cs_attackerprofilevalues.csv}
Profile,Access,Finances,Knowledge,Manpower,Motivation,Tools
Basic User,Offsite,100,2,40,Low,Low
Cybercriminal,Offsite,1000,5,160,Medium,High
Hactivist,Wireless,500,6,1500,High,Medium
Insider,Onsite,100,7,10,Medium,Medium
Nation State,Offsite,1000000,9,100000,High,High
Terrorist,Onsite,10000,4,1000,High,Medium
\end{filecontents*}
\csvautotabular{data/cs_attackerprofilevalues.csv}
	\label{table:probabilistic attacker profile}
\end{table}

\begin{table}
    \caption{Case study action profiles}
	\centering
	\begin{filecontents*}{data/cs_vulnvalues.csv}
ID,Name,Targets,A,F,K,Mp,Mo,T
V1,Remote Access Trojan,Windows 7 Machine,Offsite,0,3,20,Low,Low
V2,CVE-2017-2779,Windows 7 Machine,Onsite,10000,9,5000,High,Medium
V3,CVE-2017-2775,Windows 7 Machine,Offsite,6000,10,8000,High,Medium
V4,MODBUS MITM,NI cRIO 9064 - NI cRIO 9063,Onsite,50000,9,500,High,Medium
V5,MODBUS DOS,NI cRIO 9064 - NI cRIO 9063,Offsite,40000,6,200,High,Medium
V6,Code Injection,NI cRIO 9064,Onsite,100,4,300,Medium,Low
V7,Watering-Hole attack,Windows 7 Machine,Offsite,2000,6,300,Medium,Medium
V8,CVE-2014-4115,Windows 7 Machine,Onsite,1200,7,800,High,High
V9,CVE-2010-2568,Windows 7 Machine,Onsite,10000,8,2000,High,High
V10,CVE-2019-1713,Cisco ASA,Offsite,5000,9,100,High,Medium
\end{filecontents*}
\csvautotabular{data/cs_vulnvalues.csv}\\
	A=Access, F=Finances, K=Knowledge, MP=Manpower, M=Motivation, T=Tools
	\label{table:cs vulnerabilities values}
\end{table}

\subsection{Attacker Model Execution}
\begin{figure}
	\centering
	\includegraphics[width=7cm]{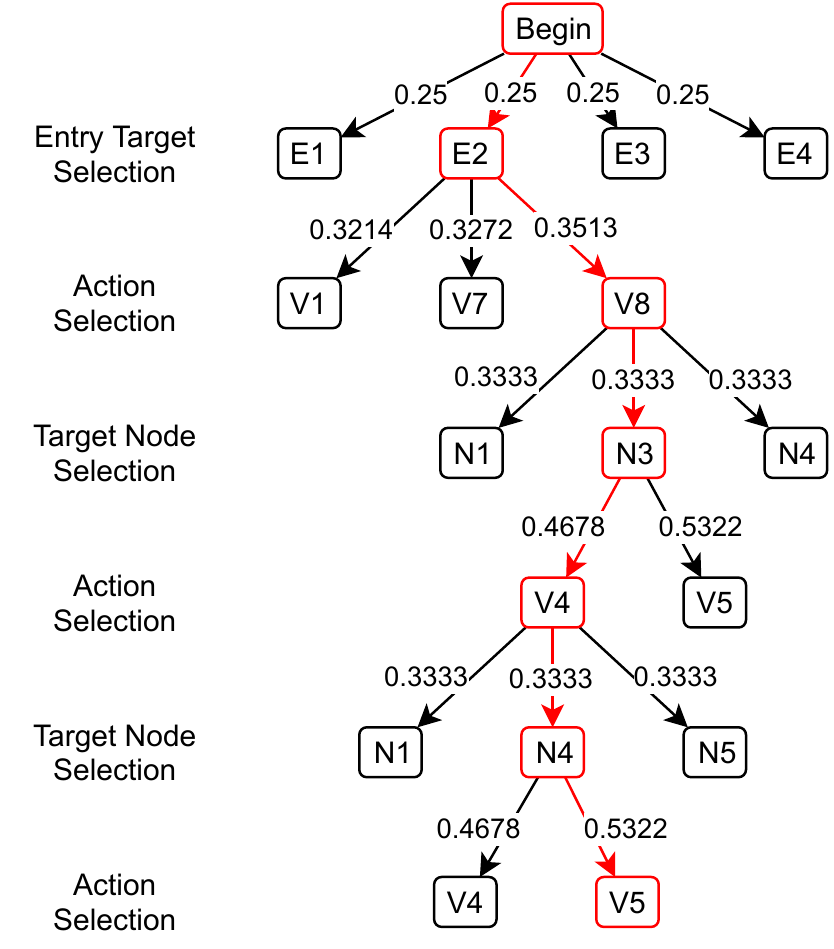}
	\caption{Attacker CPS Knowledge upon completion of the attack, including attack progression}
	\label{fig:casestudy1results}
\end{figure}

Sampling the PMF results in the selection of the Nation State attacker. The attacker profile is then evaluated against the CPS. Initial attacker knowledge assumed includes the entry vectors E1, E2, E3, and E4, along with the corresponding existence of nodes N1, N2, N6, and N7. The modeling cycle in Fig.~\ref{fig:Attacker Model Overview} begins and is repeated until either the target is reached or there are no actions remaining for the attacker to perform. Fig.~\ref{fig:casestudy1results} shows the progression of the attack as a POMDP, including each decision the attacker made in the attacker process and the probability of each decision.
\subsection{Attack Review}
The steps taken to complete the attack in Fig.~\ref{fig:casestudy1results} represent one of many possible attacks that may have been performed by the attacker. The attacker was able to compromise the CPS by exploiting three vulnerabilities. Step 1 used an infected USB thumb-drive to gain access to the monitoring workstation. Step 2 used a MODBUS man-in-the-middle attack to take over the SIS cRIO. Step 3 used a MODBUS DOS attack to disrupt the operation of the BPCS.

\section{Conclusion and Future Work} \label{sect:conclusion}
The Attacker Modeling Framework we present significantly builds upon existing research and injects a more theoretical foundation for system behavior and attacker causality models.
The flexibility of the framework readily integrates a variety of complex attacker behaviors. The proposed attack probability functions quantify the influence of attacker characteristics on the attacker's decision process and provide probabilistic predictions for the attacker behavior.
\par
Preliminary findings indicate that the proposed method scales well; specifically with respect to the decision space of a traditional MDP or attack-tree
analysis methods which would grow exponentially. The POMDP analysis framework provides manageable attack scenarios describing system vulnerabilities and putting the attack process in the context of the CPS. The case study shows the benefit of the attacker decision-by-decision analysis, allowing the cyber analyst and system engineers to have deeper insights into potential vulnerability pathways into and through the CPS.
\par
%Proposed future work includes 1) the implementation of additional rules defined in related literature, 2) the development of a tool to aid in attack scenario design and automation of the attack analysis process, 3) formalization of the AMF training and calibration process and the application of the AMF to recorded attacker data-sets from behavioral analysis studies, and 4) the integration of a penetration-testing framework such as Metasploit into the attacker-model for analysis against a SCADA test-bed.

Proposed future work includes,
\begin{enumerate}
   \item The implementation of additional rules defined in related literature
   \item The development of a tool to aid in attack scenario design and automation of the attack analysis process
   \item Exploration of different techniques to train and calibrate the AMF
   \item The application of the AMF to recorded attacker data-sets from behavioral analysis studies
   \item The integration of a penetration-testing framework such as Metasploit into the attacker-model for analysis against a SCADA test-bed
   
\end{enumerate}

\section*{Acknowledgement}
This research was made possible by NPRP 9-005-1-002 grant from the Qatar National Research Fund (a member of The Qatar Foundation). The statements made herein are solely the responsibility of the authors.

\bibliographystyle{splncs04}
\bibliography{references.bib}
\end{document}